%% file: main.tex
\RequirePackage[2020-02-02]{latexrelease}
\documentclass[runningheads]{llncs}
\usepackage{silence}
\usepackage{import}
\usepackage{preamble}
\usepackage{algorithm}
\usepackage{algpseudocode}
\usepackage{booktabs}
\usepackage[cal=boondox]{mathalfa}
\usepackage{ stmaryrd }
\usepackage{adjustbox}
\let\svthefootnote\thefootnote
\newcommand\freefootnote[1]{
  \let\thefootnote\relax
  \footnotetext{#1}
  \let\thefootnote\svthefootnote
}
\usepackage{array}
\usepackage{multirow}
\newcolumntype{P}[1]{>{\centering\arraybackslash}p{#1}}

\algnewcommand\algorithmicforeach{\textbf{for each}}
\algdef{S}[FOR]{ForEach}[1]{\algorithmicforeach\ #1\ \algorithmicdo}
\algrenewcommand\algorithmicrequire{\textbf{Input:}}
\algrenewcommand\algorithmicensure{\textbf{Output:}}
\begin{document}
%
%
\title{BeLightRec: A lightweight recommender system enhanced with BERT}
%
%

\author{Manh Mai Van\inst{1}\orcidID{0009-0005-9622-1223} \and
Tin T. Tran\inst{2}\orcidID{0000-0003-4252-6898} }
\authorrunning{Manh Mai Van and Tin T. Tran}
\titlerunning{BeLightRec: a lightweight recommender system enhanced with BERT}
%
\institute{Faculty of Information Technology, Ton Duc Thang University, Ho Chi Minh city, Vietnam\\
\email{maivanmanh@tdtu.edu.vn}  \and
Artificial Intelligence Laboratory, Faculty of Information Technology, 
Ton Duc Thang University, Ho Chi Minh city, Vietnam\\
\email{trantrungtin@tdtu.edu.vn} }
\maketitle              
\freefootnote{* Corresponding author: Tin T. Tran}
  
\begin{abstract}
The trend of data mining using deep learning models on graph neural networks has proven effective in identifying object features through signal encoders and decoders, particularly in recommendation systems utilizing collaborative filtering methods. Collaborative filtering exploits similarities between users and items from historical data. However, it overlooks distinctive information, such as item names and descriptions. The semantic data of items should be further mined using models in the natural language processing field. Thus, items can be compared using text classification, similarity assessments, or identifying analogous sentence pairs. This research proposes combining two sources of item similarity signals: one from collaborative filtering and one from the semantic similarity measure between item names and descriptions. These signals are integrated into a graph convolutional neural network to optimize model weights, thereby providing accurate recommendations. Experiments are also designed to evaluate the contribution of each signal group to the recommendation results.
\end{abstract}

\import{./}{introduction.tex}

\import{./}{background.tex}
\import{./}{method.tex}
\import{./}{experiments}

\import{./}{conlusion}
%
%

\end{document}

%% file: introduction.tex
\section{Introduction}
Recommender systems play a crucial role in both theoretical research on information retrieval and practical, everyday applications. E-commerce applications utilize recommender models to present users with items that suppliers believe the users will be interested in and likely to purchase \cite{UDOKWU2023512}. Similarly, library or book-selling applications leverage user interaction histories to curate and recommend book catalogs to users. In the initial phase of research, recommender systems focused on identifying the characteristics of target users, such as age, gender, residential address, and preferences, and then selecting items that matched these characteristics to recommend to users. However, the vast amount of product information, the rapid emergence of new items, and the challenges in collecting personal information and preferences from users pose significant difficulties. The next phase in recommender systems development addressed these challenges through collaborative filtering techniques, which identify similarities between users based on the set of items they interact with. The more items two users have interacted with in common, the higher the similarity between them. Items that a user has interacted with are then recommended to other users with high similarity to the original user.

In collaborative filtering models, the characteristic information of users and items is excluded, such as book titles, movie genres, or previous user feedback. This can lead to information loss when evaluating the similarity of items. Collaborative filtering models can be implemented on graph neural networks, where users and items are represented by vertices in the graph, and their relationships are explored through high-order propagation. Moreover, this propagation process can be efficiently implemented using prominent machine learning libraries like TensorFlow, Numpy, and Pandas. Based on collaborative filtering, we can delve deeper and exploit the similarity between items to supplement the explored graph. Items similar to those a user has interacted with should be recommended to that user. Items within the same category, or having similar descriptions or names, significantly influence the user.

\begin{figure}[ht]
\centering
\includegraphics[width=.7\textwidth]{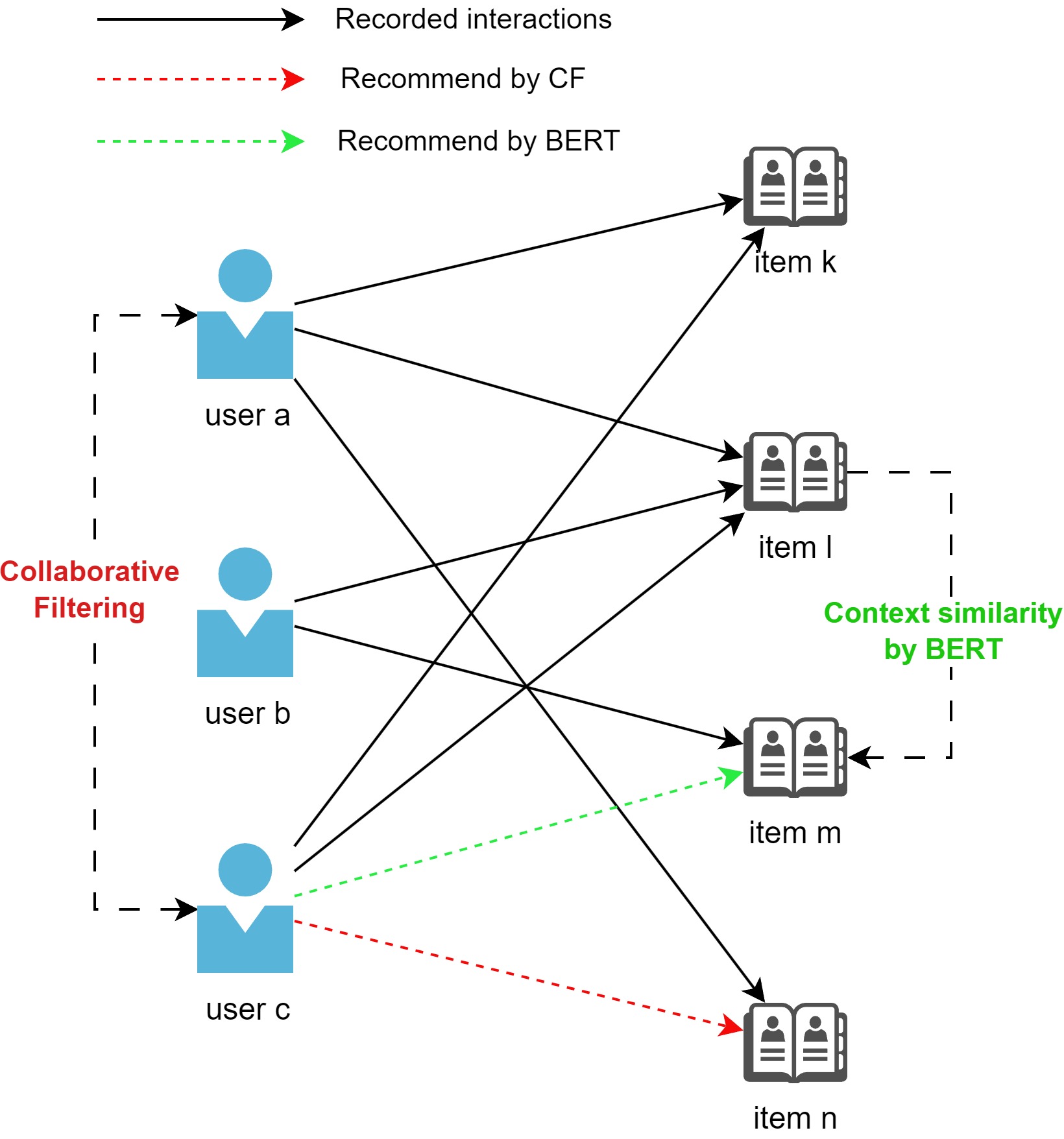}
\caption{Recommender system by collaborative filtering}
\label{fig:RSCF}
\end{figure}

With natural language processing models, evaluating the semantic similarity between two text segments can be conducted using text vectorization techniques and measuring the distance between these vectors in the language representation space. The signal propagation process on the convolutional graph network highlights the interactions between users and items, or between items themselves, through collaborative filtering, emphasizing the distinctive features of the objects. The convergence of feature values is regulated by weight matrices, which are trained with input data sets. The feature vectors of users and items are calculated after several iterations of propagation. In the following sections, we will detail the components of the recommendation system that explores the similarity between users and items, integrating semantic similarity signals between items into the output feature vectors. We summarize the recommendation model as shown in Figure \ref{fig:RSCF}. Finally, we conduct experiments on some of the most recent and real-world data sets to evaluate the proposed model.

%% file: background.tex
\section{Literature review}
\subsection{Deep learning in context-aware recommender system}
Recent developments in recommender systems, particularly since 2018, have heavily focused on leveraging deep learning and context-aware techniques to enhance personalization and accuracy. Deep learning has revolutionized the field by enabling the modeling of complex user-item interactions, surpassing traditional linear methods. Neural networks, including autoencoders, convolutional neural networks (CNNs), and recurrent neural networks (RNNs), have proven highly effective in capturing latent features and patterns from large, unstructured datasets such as user behavior, text, and multimedia content. This shift has allowed platforms like YouTube and Netflix to substantially enhance the relevance and accuracy of their recommendations. However, these models come with higher computational demands and require large-scale datasets for effective training. Research by Zhang et al. highlights the success of deep learning models in improving prediction accuracy while also acknowledging the challenges of scalability and computational cost in real-world applications \cite{10.1145/3285029}.

Additionally, context-aware recommender systems (CARS) have gained attention as they integrate contextual factors such as time, location, and user mood to enhance recommendation relevance. Contextual data, combined with deep learning models, allows for more dynamic and situation-specific recommendations, increasing user engagement and satisfaction. For example, a music recommendation system might suggest different songs depending on the time of day or a user's current activity. Publication\cite{Zhang2021} explored the impact of CARS in improving recommendation quality by adapting to real-time user environments. While these systems offer more personalized experiences, they also introduce complexity in data collection and processing, alongside concerns about user privacy and the explainability of recommendations.
\subsection{Graph Convolution Networks}
Graph Convolutional Networks (GCNs) are advanced deep learning models designed to process graph-structured data like social networks, transportation networks, and molecular structures. Unlike traditional neural networks that handle grid-like data (such as images or text), GCNs leverage the non-Euclidean structure of graphs to learn node and edge features. GCNs extend the concept of convolution from grid data to graph data, using mathematical transformations to aggregate information from neighboring nodes, as detailed in publication \cite{gcn}. This occurs through convolutional layers, where each applies a linear transformation to node features and combines them according to the graph structure \cite{WiGCN, CombiGCN}. This allows GCNs to learn complex node representations and inter-node relationships, enhancing performance in tasks such as node classification, graph classification, and link prediction. Due to their effective handling of graph-structured data and ability to uncover hidden relationships, GCNs have become a powerful and increasingly popular tool in fields such as social network analysis, computational biology, and other artificial intelligence applications.

Furthermore, Light Graph Convolution Networks (LGCN), which are models in studies \cite{ lightgcn, yu2022graph}, has shown that removing complex components such as weight matrices and bias vectors will not These increase the convergence rate of feature embeddings but also achieve better precision and recall. This is explained because the observed data sets are recorded imploit and the interaction matrix is a sparse matrix because the number of interactions of a user is very small compared to the number of all items.
 
\subsection{Natural Language Processing}
BERT (Bidirectional Encoder Representations from Transformers) \cite{bert} and TF-IDF (Term Frequency-Inverse Document Frequency) \cite{tf-idf} are prominent models in natural language processing, each serving distinct purposes. BERT, developed by Google, is an advanced language model that understands the context of words in sentences using a bidirectional approach, considering both preceding and succeeding words. It is trained on a vast amount of unlabeled text data through tasks like Masked Language Model and Next Sentence Prediction, and can be fine-tuned for various tasks such as text classification, question answering, and named entity recognition, consistently achieving superior performance in real-world applications. On the other hand, TF-IDF is used to assess the importance of a word in a document relative to a corpus of documents. It combines Term Frequency (TF), which measures a word’s frequency in a document, and Inverse Document Frequency (IDF), which measures the word’s rarity across the corpus. TF-IDF helps identify important words while reducing the impact of common words, making it useful for text search and classification. By combining the strengths of TF-IDF and BERT, applications can achieve a balance between efficiency and accuracy, leveraging TF-IDF's ability to quickly identify significant terms and BERT's deep contextual understanding to deliver high-quality results in various natural language processing tasks.

%% file: method.tex
\section{Proposed model}
We propose model BeLightRec which is based on LGCN to propagate collaborative signals and discover the feature vectors of both users and items, and enhanced with semantic similarity measures between item titles and descriptions. The model uses BERT for similarity evaluation. Furthermore, we introduce several metrics to evaluate and compare the proposed models with state-of-the-art models. 

\begin{figure}[H]
\centering
\includegraphics[width=.95\textwidth]{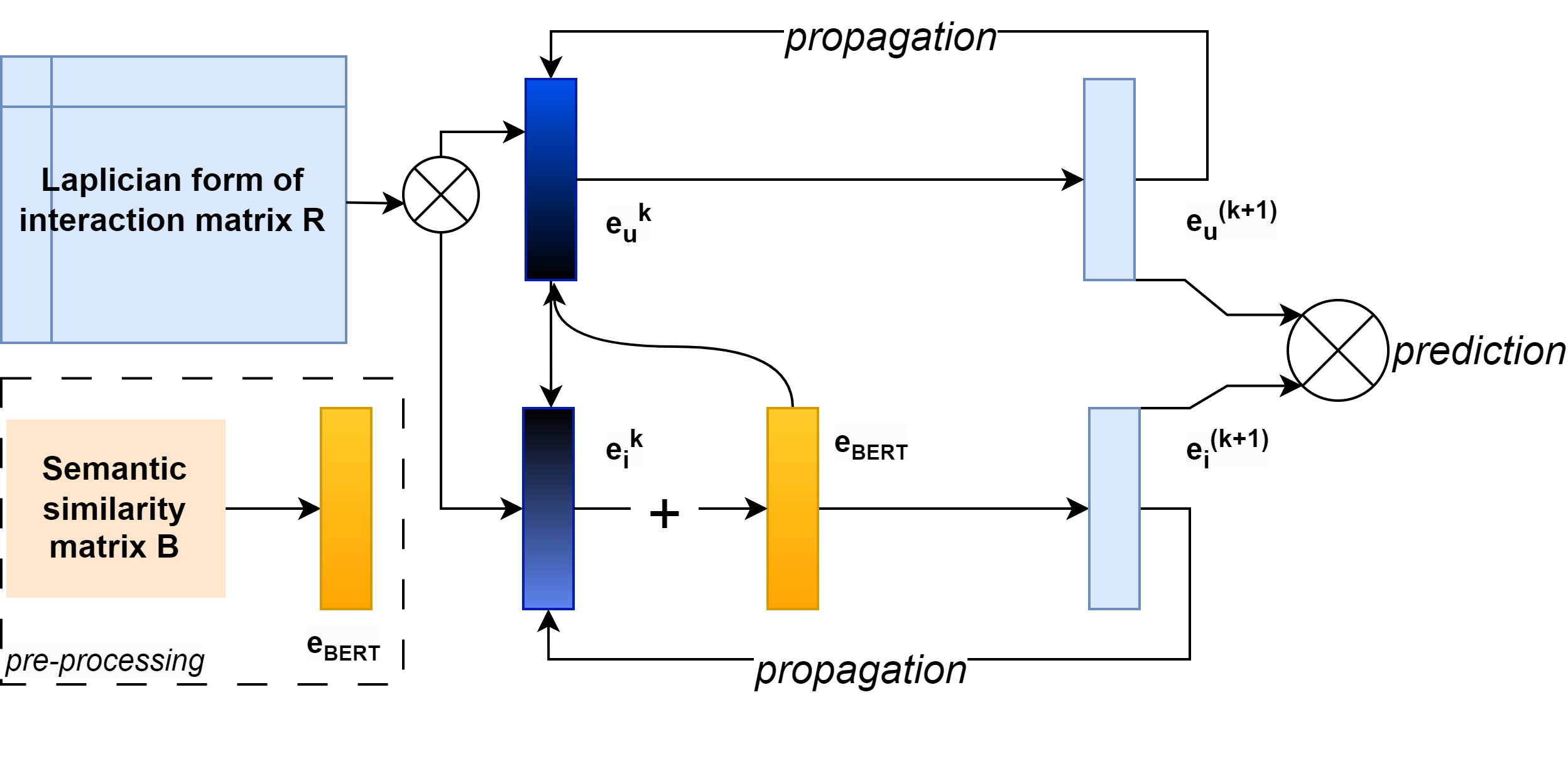}
\caption{Overview of the proposed model}
\label{fig:overview}
\end{figure}

We use two matrices as inputs to the LGCN: matrix R contains records interactions between users and items; and matrix B records semantic similarity between item titles and descriptions. All matrix elements are L1-normalized to ensure values lie between 0.0 and 1.0, minimizing the impact of different measurement units. The LGCN will propagate signals and iterate several times before finding the characteristic vector values for users and items. An overview of the model is presented in Figure \ref{fig:overview}.


\subsection{Data preprocessing}
In recommendation systems, interactions between users and items can be represented in two ways: implicit and explicit. Implicit data store interactions as binary values, while explicit data store user ratings of items during interactions. E-commerce systems often allow users to rate or provide feedback on products they have purchased or viewed. For datasets containing implicit reviews, the matrix representing user-item interactions is very sparse. Furthermore, customers with fewer than k interactions will create noise in the propagation process. In most studies, such users are removed from the dataset using k-core algorithm \cite{CombiGCN,ngcf}.

The algorithm takes a similarity matrix between users and items as input and outputs a subset of user-item pairs. It begins by calculating a ratio between the number of users and the number of items. The algorithm then examines each item to determine if it has a minimum number of interactions; items that meet this threshold are selected into a new subset. After this filtering step, the algorithm computes how many users should be chosen based on this subset of items. For each user, it identifies the items they have interacted with and calculates how closely this matches the selected subset of items using a similarity measure. The users with the highest similarity scores are selected, and the final output is generated as the set of these top users paired with the filtered items. This output represents the most relevant user-item pairs. We defined this output matrix R, and formed it out in Laplician form to serve as input to the proposed model.

\subsection{Semantic similarity measurement}
Item names and descriptions can be treated as text documents, and BERT’s tokenization technique can convert these texts into tensors. These tensors form embeddings for each initial text. The model takes the average of the token embeddings representing words in each text to create a unique vector $\overrightarrow{item_{i}}$ representation for each $item_{i}$. Finally, the model uses the cosine similarity measure to calculate the similarity between the embedding vectors of two items. The resulting score reflects the semantic similarity between two items on a 0 to 1 scale, with 1 indicating maximum similarity.
By combining BERT and TF-IDF methods, we can achieve a more accurate and comprehensive assessment of the semantic similarity between items, thereby supporting the proposed model effectively. The BERT (or TF-IDF) similarity measure between items $i$ and $j$ is described by Equation \ref{bert} to integrate both BERT and TF-IDF, the values of the similarity measures can be averaged or weighted to reflect the role of each measure.
\begin{equation}\label{bert}
BERT(item_{i},item_{j})=\cos(\overrightarrow{item_{i}},\overrightarrow{item_{j}})
\end{equation} 

Similarity matrix B: It's size is $m\times m$ where $m$ is the number of items. Each element $B_{i,j}$ represents the similarity calculated by Equation \ref{bert} based on item names and descriptions. All matrix elements are L1 normalized to ensure each value lies between 0.0 and 1.0, minimizing the impact of measurement units.

\subsection{Components of the proposed model}

\textbf{Signal propagation and embedding collection}: The feature signals of users are collected from the input matrices described in the previous section, then propagated within the model as embeddings $e_u^k$ (representing user features) and $e_i^k$ (representing item features). This process is repeated $k$ times within the LGCN \cite{lightgcn}. The embeding $\mathrm{e}_{u}^{k+1}$ at the end of each iteration is used as the input for the next propagation and is calculated by Equation \ref{embed_u}, with $\mathrm{N}_{u}^{R}$ and $\mathrm{N}_{i}^{R}$, are respectively number of neighboring users of $u$ and items of $i$ in the matrix $R$.
\begin{equation} \label{embed_u}
\mathrm{e}_{u}^{k+1} = \sum_{i\in \mathrm{N}_{u}^{R}}\frac{1}{\sqrt{\left| \mathrm{N}_{u}^{R} \right|\left| \mathrm{N}_{i}^{R} \right|}}\mathrm{e}_{i}^{k}
\end{equation} 

Simultaneously, the item feature embedding $\mathrm{e}_{i}^{k+1}$ is also calculated and propagated as shown in Equation \ref{embed_i}.

\begin{equation} \label{embed_i}
\mathrm{e}_{i}^{k+1} = \sum_{u\in \mathrm{N}_{i}^{R}}\frac{1}{\sqrt{\left| \mathrm{N}_{i}^{R} \right|\left| \mathrm{N}_{u}^{R} \right|}}\mathrm{e}_{u}^{k}+ 
\sum_{b\in \mathrm{N}_{i}^{B}}\frac{1}{\sqrt{\left| \mathrm{N}_{i}^{B} \right|\left| \mathrm{N}_{b}^{B} \right|}}\mathrm{e}_{i}^{k}
\end{equation} 

After several iterations of signal propagation, the characteristic vectors of users and items are found by Equation \ref{embed_final}.

\begin{equation} \label{embed_final}
e_{u}=\frac{1}{K}\sum_{k=1}^{K}\mathrm{e}_{u}^{k} \quad; \quad
e_{i}=\frac{1}{K}\sum_{k=1}^{K}\mathrm{e}_{i}^{k}
\end{equation} 

\textbf{Prediction and loss function}: The characteristic vectors for users and items converge after several iterations of propagation, and the prediction score between user $u_i$ and item $i_j$ can be calculated by Equation \ref{predict}.

\begin{equation} \label{predict}
\widehat{y}_{ui}={e_{u_i}^*}^\top  e^*_{i_j}
\end{equation} 

The Bayesian personalized ranking (BPR) method is the optimal choice for implementing the loss function because it is the most effective ranking method for datasets with implicit feedback \cite{bprmf}. We use two observation sets: $\Omega^+_{ui}$ representing the interacted items and $\Omega^-_{ui}$ representing the non-interacted items. The loss function is calculated by Equation \ref{loss}.
\begin{equation} \label{loss}
    Loss_{bpr} =  \sum_{\Omega^+_{ui}}\sum_{\Omega^-_{uj}} -ln \sigma ( \widehat{y}_{ui} - \widehat{y}_{uj} ) +  \lambda  \parallel  \Phi   \parallel ^2_2
\end{equation} 

%% file: experiments.tex
\section{Experiments}
\subsection{Datasets description}
We obtained the latest datasets and removed users with fewer than 10 interactions to enrich the training set. We split the data into training and testing sets with a ratio of 80\% and 20\%, respectively. Statistics on the datasets are described in Table \ref{tab:data}. In that table, we record the number of users and items, the number of interactions in the survey dataset, density is the ratio of the number of interactions to the product of the number of users with items, and average text count is the average number of characters in the description of the items in the dataset.
\begin{itemize}
           \item \textbf{Amazon-Book dataset} is widely used in the field study of recommendation systems. It is one of the datasets provided by Amazon, including detailed information about book products, user reviews, ratings, and other interactions between users and books. 
    \item \textbf{Recipes dataset} is a collection of information about various dishes and their preparation methods. These datasets might also have additional information like recipe names, cuisine types, user ratings, reviews, and images of the completed dishes. 
    \item \textbf{Hawaii dataset} is part of a larger collection of location-based data from Google Local, hosted by UCSD. This dataset focuses on providing detailed information about various locations in Hawaii, including restaurants, hotels, and attractions.  
\end{itemize}

\begin{table}
\centering
\caption{Statistic of the experiment datasets.}
\label{tab:data}
\begin{tabular}{|l|r|r|r|r| r|}
\hline
\textbf{Dataset} &  \textbf{\#Users} & \textbf{\#Items} & \textbf{\#Interactions} & \textbf{Density} & \textbf{Ave. text}\\
\hline
AmazonBook& 27,034 & 32,157 & 992,700 & 1.14e-3 & 8,599\\
Recipes& 3,969 & 5,637 & 188,135 & 8.41e-3 & 344\\
Hawaii& 35,604 & 9,111 & 1,135,062 & 3.5e-3 & 102\\
\hline
\end{tabular}
\end{table}

\subsection{Baseline models} We use the same datasets and repeat the experiments on all the following baseline models to demonstrate the result:
\begin{itemize}
    \item \textbf{BPR-MF} \cite{bprmf} is a popular method in recommendation systems, combining Matrix Factorization (MF) and Bayesian Personalized Ranking (BPR). MF decomposes the user-item rating matrix into two smaller matrices representing users and items to predict ratings. BPR optimizes personalized ranking by maximizing the probability that an interacted item is ranked higher than a non-interacted one.
    \item \textbf{NGCF} \cite{ngcf} leverages graph structures to capture complex interaction information. Users and items are represented as nodes in a graph, with interactions as connecting edges. NGCF applies graph neural network layers to propagate and aggregate information across layers, enabling the model to learn high-order relation features.
    \item \textbf{LightGCN} \cite{lightgcn} simplifies traditional graph neural networks by removing complex components like transformation matrices and non-linear activation functions, focusing instead on the aggregation and propagation of information between nodes.
    \item  \textbf{BERT} only uses semantic similarity measure between items and does not have signal propagation loop like other GCN models.
    \item \textbf{BeLightRec} is the LGCN model enhanced with BERT. The semantic similarity signal between items should be collected during the propagation process.
    \item \textbf{BeLightRec+W} is a traditional GCN with weight matrices and bias vector. We implemented this model as a variant of proposed LGCN.
\end{itemize}

\subsection{Experimental settings and metrics}
Precision, recall, and NDCG@k are key metrics used to evaluate the performance of information retrieval and recommendation systems.
\begin{itemize}
\item \textbf{Precision} measures the proportion of relevant items retrieved out of the total items retrieved. It reflects the accuracy of the system in returning relevant results. High precision means that the results returned are mostly relevant.

\item \textbf{Recall} assesses the proportion of relevant items retrieved out of the total relevant items available. It indicates the system's ability to find all relevant items. High recall means that the system can identify most of the relevant items.

\item \textbf{NDCG@k} (Normalized Discounted Cumulative Gain at k) is a comprehensive metric that considers the relevance and the position of the retrieved items. It is particularly useful in ranking tasks. NDCG@k evaluates the quality of the top-k results by accounting for the graded relevance of items, giving higher scores to more relevant items appearing higher in the list. We conducted experiments with Top-5 and Top-20 settings in our experiments.
\end{itemize}
To ensure fair experimental results, we maintain consistent parameters across all models. Specifically, we set the learning rate to 0.001, the L2 normalization coefficient to $1 \times10^{-5}$, and the number of LGCN layers to three, with each layer having an embedding size of 64. We also apply the same early stopping strategy used by NGCF and LightGCN.

\subsection{Experiment results}
\begin{table}
\centering
\caption{Overall performance comparisons}\label{tabResult}
\begin{adjustbox}{width=\columnwidth,center}
\begin{tabular}{| c|c| c|c|c| c|c|c| c|c|}
\hline
 \textbf{Dataset} & \multicolumn{3}{c|}{\textbf{AmazonBook}} & \multicolumn{3}{c|}{\textbf{Recipes}} & \multicolumn{3}{c|}{\textbf{Hawaii}}\\
   & recall & precision & ndcg &  recall & precision & ndcg & recall & precision & ndcg \\ 
\hline   
\multicolumn{10}{l}{\textit{Top-5 results}}\\
\hline
BPR-MF       & 0.01965 & 0.01968 & 0.02382 & 0.01490 & 0.02354 & 0.02580 & 0.04656 & 0.05531 & 0.06341 \\
NGCF         & 0.01856 & 0.01912 & 0.02144 & 0.01612 & 0.02526 & 0.02631 & 0.05210 & 0.06235 & 0.07136\\
LightGCN     & 0.02074 & 0.02133 & 0.02374 & 0.01624 & 0.02440 & 0.02625 & 0.05368 & 0.06398 & 0.07343\\
BeLightRec+W & 0.02063 & 0.02117 & 0.02375 & 0.01653 & 0.02564 & 0.02749 & 0.05441 & 0.06479 & 0.07431\\
BERT   & 0.02228 & 0.02265 & 0.02528 &  0.01534 &  0.02358  &  0.02590  & 0.05340 &  0.06405 & 0.07308\\
BeLightRec   & \textbf{0.02610} & \textbf{0.02653} & \textbf{0.02938} & \textbf{0.01789} & \textbf{0.02681} & \textbf{0.02944} & \textbf{0.05782} & \textbf{0.06896} & \textbf{0.07935}\\
\hline
\multicolumn{10}{l}{\textit{Top-20 results}}\\
\hline    
BPR-MF       & 0.56010 & 0.01492 & 0.03948 & 0.04499 & 0.01848 & 0.03567 & 0.11835 & 0.03540 & 0.08817 \\
NGCF         & 0.05502 & 0.01490 & 0.03664 & 0.05090 & 0.01988 & 0.03790 & 0.13709 & 0.04174 & 0.10134\\
LightGCN     & 0.05851 & 0.01582 & 0.03940 & 0.05186 & 0.02012 & 0.03875 & 0.13741 & 0.04179 & 0.10264\\
BeLightRec+W & 0.05886 & 0.01603 & 0.03971 & 0.05180 & 0.02079 & 0.03911 & 0.14026 & 0.04265 & 0.10441\\
BERT  & 0.06720 & 0.01811 & 0.04424 & 0.04681 & 0.01898 & 0.03659 & 0.13943 & 0.04252 & 0.10323\\
BeLightRec   & \textbf{0.07563} & \textbf{0.02034} & \textbf{0.05020} & \textbf{0.05415} & \textbf{0.02125} & \textbf{0.04148} & \textbf{0.14628} & \textbf{0.04447} & \textbf{0.10991}\\
\hline
\end{tabular}
\end{adjustbox}
\label{figresult}
\end{table}
The overall performance comparison is shown in Table \ref{tabResult} with all models on three datasets. The results showed that BeLightRec consistently outperforms other models across all datasets and metrics. Even though BPR-MF is not a GCN-based models, we still experiment with that model as a classic benchmark. The results progressively improve from BPR-MF to NGCF and LightGCN, aligning with published insights on these models. The LightGCN and BeLightRec models both outperform their corresponding models, NGCF and BeLightRec+W. This demonstrates the effectiveness of LGCN on implicit datasets.

We split the results table into two parts for Top-5 and Top-20. In each part, the AmazonBook, Recipes, and Hawaii datasets are presented respectively. In each dataset, the three metrics recall, precision, and ndcg are applied to each of the mentioned models.

\subsection{Ablation study}
\subsubsection{Contribution of each propagation signal}
During propagation, the model aggregates the similarity signal between items using an interactive filtering method with the semantic similarity signal using the BERT model's measurement. To show the contribution of each signal, we removed one of the two sources in turn and compared the increase in precision, recall, and NDCG@5 in the conducted experiment. The LightGCN model does not care about the semantic similarity signal measured by BERT, while the BERT model in our experiment does not repeat the propagation process of GCN.

\begin{figure}[H]
\centering
\includegraphics[width=.85\textwidth]{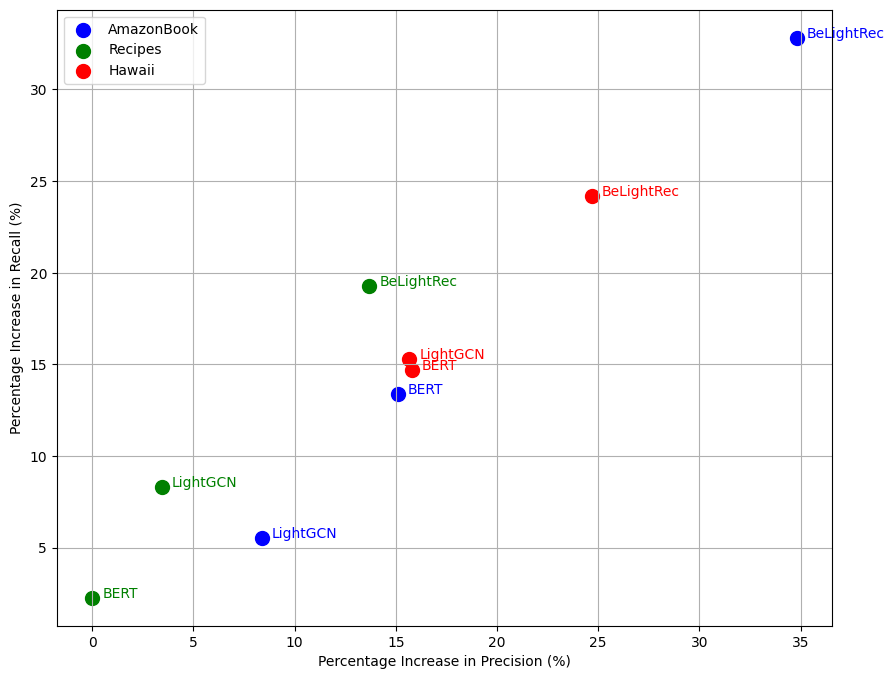}
\caption{Percentage increase in Precision vs Recall from BPR-MF}
\label{fig:incper}
\end{figure}

Based on the experimental results, we depict them in Figure \ref{fig:incper} to show the percentage improvement of each model compared to the classic MF-BPR model. We find that there are 3 comparison scenarios between the two signal sources under discussion as follows

\begin{itemize}
    \item \textbf{CF is better than BERT in dataset Recipes}: In the recipe descriptions, the text describing each item is presented as a list of ingredients with very little specific description of them. Users write about the recipes in terms of how they feel rather than the objective description of the recipe. This makes it difficult for the BERT model to capture semantic similarity signals and contributes almost nothing to the precision metric when applied alone.
    \item \textbf{BERT is better than CF in dataset AmazonBook}: The book titles and their reviews are very detailed and directly related to the book content in this dataset. This has led to the effectiveness of the BERT model in assessing semantic similarity between items, i.e. books.
    \item \textbf{CF and BERT are equal in dataset Hawaii}: This is an experiment where the two sources of signal contribute roughly equally. First, although the item descriptions in the Hawaii dataset are shorter in length than the item descriptions in the Recipes dataset, they focus on describing locations and geographic information. Second, the correlation between users measured based on the locations they visited suggests that users chose items based on category rather than description.
\end{itemize}

In all three comparison scenarios, when combining both signal sources, the results lead to a sharp increase in the values of precision and recall measures. This has demonstrated the superiority of the GCN model and the addition of the semantic similarity signal source between items in each propagation step.

\subsubsection{The effectiveness of item description text presentation}

\begin{table}
\centering
\caption{Text descriptions of items in datasets}\label{tabExample}
\begin{tabular}{| c|c| p{0.75\linewidth}|}
\hline
\textbf{Dataset} & \textbf{Item} & \textbf{Item description} \\
\hline
AmazonBook & \#60 & Russian Winter: A Novel Daphne Kalotay (Author) A mysterious jewel holds the key to a life-changing secret, in this breathtaking tale of love and art, betrayal and redemption. When she decides to auction her remarkable jewelry collection, Nina Revskaya, once a great star of the Bolshoi Ballet, believes she has finally drawn a curtain on her past. Instead, the former ballerina finds herself overwhelmed by memories of her homeland and of the events, both glorious and heartbreaking, that changed the course of her life half a century ago.\\ \hline
 AmazonBook & \#3171 & Blaze: A Novel Richard Bachman: From Publishers Weekly Written circa 1973, this trunk novel, as Bachman's double (aka Stephen King) refers to it in his self-deprecating foreword, lacks the drama and intensity of Carrie and the horror opuses that followed it. Still, this fifth Bachman book (after 1996's The Regulators ) shows King fine-tuning his skill at making memorable characters out of simple salt-of-the-earth types. Clayton Blaze Blaisdell has fallen into a life of delinquency ever since his father's brutal abuse rendered him feebleminded. \\ \hline
Recipes & \#2693 & 15 minute shrimp scampi - truly delicious seafood/pasta dish that is nearly impossible to mess up \& a great stand by for rushed dinners. ['start salted water boiling for pasta', 'heat olive oil', 'saute garlic \& onions over high heat till onions start to turn clear', 'add wine , turn to medium heat \& reduce for 5 minutes add shrimp till warmed through', 'stir in butter till melted', 'turn heat to lo / warm', 'salt \& pepper to taste', 'cover till pasta is ready', 'serve together']\\ \hline
Recipes & \#3644 & smoky refried beans - 'leave some whole for texture' "['heat oil in a heavy skillet over medium-low heat', 'add whole garlic cloves and cook , turning once , until browned on both sides , about 5 minutes', 'smash garlic cloves with a fork just enough to break them open and flatten them a little', 'add pintos and liquid and cook until heated through , about 5 minutes', 'add spices and salt to taste', 'stir well', 'turn the heat down to medium-low and smash beans with a big fork', ""you don't have to smash every single bean""\\ \hline
Hawaii & \#1821  & McDonald's Fast food restaurant Breakfast restaurant Coffee shop Hamburger restaurant Restaurant Sandwich shop - Classic, long-running fast-food chain known for its burgers, fries \& shakes.\\ \hline
Hawaii & \#7843  & Quiksilver Clothing store Men's clothing store Skateboard shop Surf shop Women's clothing store \\

\hline
\end{tabular}
\end{table}

We randomly select a few item descriptions from the three datasets in the experiment, and list them in Table \ref{tabExample}, to observe the grammatical and lexical structure of those descriptions. We found that the item descriptions of each dataset were presented as follows.
\begin{itemize}
    \item \textbf{AmazonBook}: The items in this dataset are books, and they have very detailed and clear descriptions. Each description starts with the book title, author name, summary, and the reviewer’s sentiment. For that reason, the BERT model does a good job of assessing the semantic similarity between pairs of items, and this contributes to the output of our proposed model.
    \item \textbf{Recipes}: The item descriptions in this dataset start with the dish name, followed by a list of ingredients in an array, and finally the steps to prepare it. Despite having longer text lengths than the item descriptions in the Hawaii dataset, the semantic similarity measure in the Recipes dataset did not perform well when used as a single signal. We also assume that ingredient similarity does not necessarily lead to dish similarity.
    \item \textbf{Hawaii}: Although the item descriptions in this dataset are the shortest in length, they contain a lot of keywords. These keywords often refer to the classification and properties of the places, i.e., the items. This factor is what makes the semantic similarity measure in the Hawaii dataset not worse than the AmazonBook dataset, even though the description text length of the Hawaii dataset is the shortest among the experimental datasets.
\end{itemize}

It is clear that the grammatical structure, vocabulary and writing style in the item descriptions of the datasets contribute greatly to the semantic similarity measure measured by BERT.

%% file: conlusion.tex
\section{Conclusion}
We have presented a machine learning model utilizing graph convolutional networks to explore the similarities between items based on their names and descriptions using metrics from the BERT and TF-IDF language models. Applying natural language processing models to a graph neural network-based data mining model is challenging but has yielded better results by uncovering item information that collaborative filtering methods missed. Our model consists of multiple modules that allow for enhancements or customizations when applied to specific recommendation systems.

The semantic correlation between items measured by BERT model has contributed significantly to the models' metrics. Therefore, further in-depth research on semantics should be conducted in the future. First, the data pre-processing step should normalize text segments and remove meaningless characters. Second, the distance between each pair of vectorized item descriptions needs to be improved in terms of calculation formula. Finally, the models should adjust the influence weight of semantic similarity measure based on the nature of the item descriptions.

%% file: main.bbl
\begin{thebibliography}{99}

\bibitem{UDOKWU2023512}Udokwu, C., Zimmermann, R., Darbanian, F., Obinwanne, T. \& Brandtner, P. Design and Implementation of a Product Recommendation System with Association and Clustering Algorithms. {\em Procedia Computer Science}. \textbf{219} pp. 512-520 (2023), https://doi.org/10.1016/j.procs.2023.01.319
\bibitem{10.1145/3285029}Zhang, S., Yao, L., Sun, A. \& Tay, Y. Deep Learning Based Recommender System: A Survey and New Perspectives. {\em ACM Comput. Surv.}. \textbf{52} (2019,2), https://doi.org/10.1145/3285029
\bibitem{Zhang2021}Zhang, L., Li, X., Li, W., Zhou, H. \& Bai, Q. Context-Aware Recommendation System using Graph-based Behaviours Analysis. {\em Journal Of Systems Science And Systems Engineering}. \textbf{30}, 482-494 (2021,8,1), https://doi.org/10.1007/s11518-021-5499-z
\bibitem{gcn}Kipf, T. \& Welling, M. Semi-Supervised Classification with Graph Convolutional Networks.  (2017), https://doi.org/10.48550/arXiv.1609.02907
\bibitem{WiGCN}Tran, T. \& Snasel, V. Improvement Graph Convolution Collaborative Filtering with Weighted Addition Input. {\em Intelligent Information And Database Systems}. pp. 635-647 (2022), https://doi.org/10.1007/978-3-031-21743-2\_51
\bibitem{CombiGCN}Nguyen, L. \& Tran, T. CombiGCN: An Effective GCN Model for Recommender System. {\em Computational Data And Social Networks}. pp. 111-119 (2024), https://doi.org/10.1007/978-981-97-0669-3\_11
\bibitem{lightgcn}He, X., Deng, K., Wang, X., Li, Y., Zhang, Y. \& Wang, M. LightGCN: Simplifying and Powering Graph Convolution Network for Recommendation. {\em Proceedings Of The 43rd International ACM SIGIR Conference On Research And Development In Information Retrieval}. pp. 639-648 (2020), https://doi.org/10.1145/3397271.3401063
\bibitem{yu2022graph}Yu, J., Yin, H., Xia, X., Chen, T., Cui, L. \& Nguyen, Q. Are Graph Augmentations Necessary? Simple Graph Contrastive Learning for Recommendation. (2022), https://doi.org/10.1145/3477495.3531937
\bibitem{bert}Devlin, J., Chang, M., Lee, K. \& Toutanova, K. BERT: Pre-training of Deep Bidirectional Transformers for Language Understanding. (2019), https://doi.org/10.48550/arXiv.1810.04805
\bibitem{tf-idf}Bafna, P., Pramod, D. \& Vaidya, A. Document clustering: TF-IDF approach.  (2016,3), https://doi.org/10.1109/ICEEOT.2016.7754750
\bibitem{bprmf}Rendle, S., Freudenthaler, C., Gantner, Z. \& Schmidt-Thieme, L. BPR: Bayesian Personalized Ranking from Implicit Feedback. {\em Proceedings Of The Twenty-Fifth Conference On Uncertainty In Artificial Intelligence}. pp. 452-461 (2009), https://doi.org/10.48550/arXiv.1205.2618
\bibitem{ngcf}Wang, X., He, X., Wang, M., Feng, F. \& Chua, T. Neural Graph Collaborative Filtering. {\em Proceedings Of The 42nd International ACM SIGIR Conference On Research And Development In Information Retrieval}. (2019,7), https://doi.org/10.1145/3331184.3331267


\end{thebibliography}
